\documentclass[11pt,twoside]{article}
\usepackage{asp2010}

\resetcounters

\begin{document}

\title{Thermal and Non-thermal radiation from pulsars: hints of physics}
\author{Shi Dai and Renxin Xu
\affil{School of Physics and State Key Laboratory of Nuclear Physics
and Technology, Peking University, Beijing, 100871,
China;\\
{\tt Email:} \{daishi, r.x.xu\}@pku.edu.cn}}

\begin{abstract}
Thermal and non-thermal radiation from pulsars carries significant
information from surface and would have profound implications on the
state of dense matter in compact stars. For the non-thermal radio
emission, subpulse drifting phenomena suggest the existence of
Ruderman-Sutherland-like gap-sparking and strong binding of
particles on pulsar polar caps. While conventional neutron star
models can hardly provide such a high binding energy, the strong
self-bound surface of quark-cluster stars can naturally solve this
problem. As for the thermal one, the featureless X-ray spectra of
pulsars may indicate a bare surface without atmosphere, and the
ultrarelativistic fireball of $\gamma$-ray bursts and supernovae
would also require strong self-bound surfaces.
Recent achievements in measuring pulsar mass and mass-radius
relation further indicate a stiff equation of state and a
self-bound surface.
Therefore, we conjecture that matters inside pulsar-like compact
stars could be in a quark-cluster phase. The surface of quark-cluster
stars is chromatically confined and could initially be bare. Such a
surface can not only explain above features, but may also promote a
successful core-collapse supernova, and the hydro-cyclotron oscillation
of the electron sea above the surface could be responsible for those
absorption features detected in the X-ray spectrum.

\end{abstract}

It was meaningful to celebrate Janusz' 60th birthday on the occasion of the
80th anniversary that Landau speculated dense matter at nuclear density inside stars.
The speaker (RX) is focusing on pulsars from the dense matter point
of view.
Janusz has made contributions to pulsar study, in both observation
and modeling. Certainly we had effective discussions and a close
relationship, just as the fact that pulsar magnetospheric and inner
researches are strongly coupled. Wish Janusz a happy and healthy
life!

\section{Non-thermal radiation: bound strongly on surface?}

Radio radiation is the most important non-thermal radiation from pulsars, and it seems
that all radio subpulses are drifting.
In~\citet{SR75}, a vacuum gap above the polar cap of a pulsar was first
suggested. In this scenario, the sparks produced by the inner-gap breakdown result in the
sub-pulses, and the $E \times B$ drift causes the observed drifting features. However, in
order to form such a vacuum gap, the binding energy of particles on pulsar polar caps must
be very high, and calculations have shown that the binding energy of Fe at the neutron
star surface is not enough~\citep{Flowers77,Lai01}.
One way to solve this binding energy problem could be the partially
screened inner gap model in the regime of conventional neutron
star~\citep{Gil03,Gil06}, but an alternative way would be in the bare
quark-cluster star model~\citep{Yu11}.

Quarks on the surface of quark-cluster stars are chromatic confined, so the binding energy
of quarks can be considered as infinity compared to electromagnetic interaction. As for
electrons on the surface, the binding energy is determined by the difference between the height
of the potential barrier in the vacuum gap and the total energy of electrons.
In~\citet{Yu11}, the magnetospheric activity of bare quark-cluster star was investigated
in quantitative details, and they have shown that the huge potential barrier built by
the electric field in the vacuum gap above the polar cap can usually prevent electrons
from streaming into the magnetosphere, unless the electric potential of a pulsar is
sufficiently lower than that at the infinite interstellar medium.
Therefore, both positive and negative particles on the surface of
quark-cluster stars are bound strongly, and a vacuum gap above the polar
cap could be naturally formed to reproduce the drifting subpulses.

\section{Thermal radiation: featureless spectrum and clear fireball?}

The thermal spectra of pulsars differ substantially from Planck-like
ones in conventional neutron star models~\citep{Zav96}. Many
calculations~\citep[e.g.][]{Roma87,Zav96} also showed that spectral
lines should be detectable with the spectrographs on board Chandra
and XMM-Newton. However, up to now, no atomic line has been observed
with certainty, and detailed spectral analysis of the combined X-ray
and optical data of RX J1856.5$-$3754 have shown that no atmosphere
model can fit the data well~\citep{Bur01}.
These results may suggest a bare surface of pulsars.

Nevertheless, the best absorption features were detected for the
central compact object 1E 1207.4$-$5209 at $\sim0.7$ keV and
$\sim1.4$ keV~\citep{San02,Mere02,Big03}. These lines might be of
electron-cyclotron origin~\citep{Gott07,Hal11}, but such a simple
single particle approximation might not be reliable due to the high
electron density on strange stars. \citet{Xu12} investigated the
global motion of the electron sea on the magnetized surface of
quark-cluster stars. Their calculations showed that both the
frequency and strength of the absorption features can be understood
by the hydrocyclotron oscillations of the electron sea. This
mechanism may also explain the detected lines in the burst spectrum
of SGR 1806$-$20 and those in other dead pulsars~\citep{Xu12}.

The ultrarelativistic fireball of $\gamma$-ray bursts and supernovae
also requires a bare and strong self-bound surface, since the total
mass of baryons can not be too high, otherwise baryons would carry
out too much energy of the central engine. However, the number of
baryons loaded with the fireball is unlikely to be small for
conventional neutron stars, considering their weak-bound surface and
the extremely high luminosity of the fireball. This baryon contamination
problem can be solved if the central compact objects are
strange stars, because baryons are confined by strong color
interaction while $e^\pm$-pairs, photons, neutrino pairs and
magnetic fields can escape from the surface~\citep{Paczy05,Cheng96}.
Such a self-bound surface may also promote core-collapse supernovae.
A nascent quark-cluster star born in the center of GRB or supernova
would radiate Planck-like thermal emission due to its ultrahigh
surface temperature, and the photon luminosity is not constrained by
the Eddington limit since the surface of quark-cluster stars could
be bare and is chromatic confined. Enormous thermal emissions could
provide strong radiation pressure and are promising to alleviate the
current difficulty in core-collapse supernovae~\citep{Chen07}.

\section{Hints from surface: a quark-cluster state in the QCD phase diagram?}

It is challenging to understand the states in the QCD phase diagram,
especially at high density.
Conventional neutron and quark stars are models for us to understand
the inner structure of pulsars. For conventional neutron stars, the
highly non-perturbative strong interaction makes quarks grouped into
neutrons, while for quark stars, the perturbative strong interaction
inside makes quarks to be almost free if the coupling is weak.
However, quarks in dense matter at realistic baryon densities
($\rho\sim2-10\rho_{\rm{0}}$) could be coupled strongly and grouped
into quark clusters, as relativistic heavy ion collision experiments
have shown that the interaction between quarks is very strong even
in hot quark-gluon plasma~\citep{Shur09}. We conjecture that matters
inside pulsars could be in a quark-cluster phase, and the star could
be in a solid state since the kinetic energy of quark clusters
should be much lower than the interaction energy between the
clusters.

Such solid quark-cluster stars are different from conventional neutron stars in two aspects~\citep{Xu11}:
(1) The equation of state determines the global structure: liquid or rigid? soft or stiff?
(2) The surface is gravity-bound for neutron stars while it is self-confined by strong force
for quark-cluster stars.
The global solid state of quark-cluster stars can explain the observation of possible precessions
of pulsars~\citep{Sta00}. The latent heat released during the phase transition of quark-cluster
stars from liquid to solid may provide a long-term steady central engine, and interpret the long-lived
plateau in some GRB afterglows~\citep{Dai11}. The huge gravitational and elastic energy released
during star quarks may also power the flares and bursts of soft $\gamma$-ray repeaters and
anomalous X-ray pulsars~\citep{Xu07}.

The peculiar surface properties of quark-cluster stars would result in different manifestations
from that of conventional neutron stars and may indicate the state of dense matter inside
pulsars. On one hand, the bare and chromatic confined surface can naturally explain the
drifting subpulses and featureless thermal spectra of pulsars, and provide clean fireball for
supernovae and $\gamma$-ray bursts as we discussed above. On the other hand, the mass-radius (M-R)
relation of quark-cluster stars would be different from that of conventional neutron stars,
because self-bound quark-cluster stars have non-zero surface density, and their radii usually
increase as masses increase. Additionally, the state equation of quark-cluster stars is stiff
because of the non-relativistic motion of quark clusters, and then a much larger maximum mass
is expected.
Recent mass and radius determination of the Rapid Burster (MXB
1730-335)~\citep{Sala12} and the two solar mass neutron
star~\citep{Dem10} have set strict constrains on the M-R relation
and maximum mass of pulsars. Because of the hyperon puzzle (Haensel,
this proceedings) and possible quark-deconfinement puzzle due to
asymptotic freedom, most conventional neutron star and quark star
models are hard to satisfy these two constrains, but quark-cluster
stars can pass these tests~\citep{LaiX09,Na11}.
Both global and surface properties of quark-cluster stars are quite different from that of
conventional neutron stars. Future observations and studies are hopeful to distinguish them,
and then promote our understanding of the state of dense matter and the strong interaction.

\section{Conclusions}

Diverse manifestations of thermal and non-thermal radiation from
pulsars hint the surface properties and the state of compact stars.
The drifting subpulses of non-thermal radio radiation suggest the
strong binding of particles on pulsar polar caps, and the
featureless thermal spectra and the ultrarelativistic fireball of
$\gamma$-ray bursts and supernovae indicate that pulsars may be
strong self-bound and bare without atmosphere.

The bare and chromatic confined surface of quark-cluster stars can
not only explain the drifting subpulses and featureless thermal
spectra, but also promote core-collapse supernovae and understand
the absorption features by the hydrocyclotron oscillations of the
electron seas. What is more, the self-bound surface and the stiff
equation of state of quark-cluster stars can fit the measurement of
mass and radius of compact stars.

\acknowledgements We would like to thank useful discussions at the
pulsar group of PKU. This work is supported by the National Natural
Science Foundation of China (Grant Nos. 10935001, 10973002), the
National Basic Research Program of China (Grant Nos. 2009CB824800,
2012CB821800), and the John Templeton Foundation.

\bibliography{aspauthor}

\end{document}